\documentclass[twocolumn,showpacs,preprintnumbers,amsmath,amssymb]{revtex4}
\usepackage [dvips]{graphicx}

\begin{document}

\title {On the current gap of single-electron transistors}

\author {Valentin V. Pogosov\footnote {Corresponding author:
vpogosov@zntu.edu.ua}, Eugene V. Vasyutin}

\address {Department of Microelectronics and Semiconductor Devices,
Zaporozhye National Technical University, Zhukovsky Str. 63,
Zaporozhye 69064, Ukraine}

\date {\today}

\begin {abstract}
Effects of the single-electron tunneling and the Coulomb blockade in
a cluster structure (the molecular transistor) are investigated
theoretically. In the framework of the particle-in-a-box model for
the spherical and disk-shaped gold clusters, the electron spectrum,
the temperature dependence of the chemical potential and the
residual charge are calculated. We show that the residual charge is
equal to the non-integer value of elementary charge $e$ and depends
on the cluster's shape. The equations for the analysis of the
current-voltage characteristic are used under conservation condition
for the total energy of the structure taking into account the
contact potential difference. Restrictions associated with the
Coulomb instability of a cluster are introduced into the theory in a
the simple way. It is shown that the critical charge of the cluster
in the open electron system is close to the residual charge. For
single-electron transistors based on small gold clusters the current
gap and its voltage asymmetry  are computed. We demonstrate that the
current gap exhibits non-monotonic size dependences, which are
related to the quantization of the electron spectrum and the Coulomb
blockade.
\end {abstract}



\maketitle

\section {Introduction}

Metal granules, which are weakly coupled via tunnel barriers to
electron reservoirs, are of considerable interest in physics of
low-dimensional systems (see reviews \cite {17., Ralph} and
references therein).

The tunneling current flowing through two massive electrodes can be
controlled, if a cluster is placed between them. At first sight, the
probability of electron tunneling (and consequently a value of the
current) should be much greater in the presence of a granule between
the reservoirs, than in the case of its absence. However, an
opposite behavior was observer in experiments for the spherical-like
\cite {Lih, Ohgi-2001, Ohgi, Ohgi-2004} and disk--shaped \cite
{Wang, Hou} clusters. Measured $I-V$ characteristics have a plateau
of the zero current (a current gap).

\begin {figure} [! t! b! p]
\centering
\includegraphics [width =.46\textwidth] {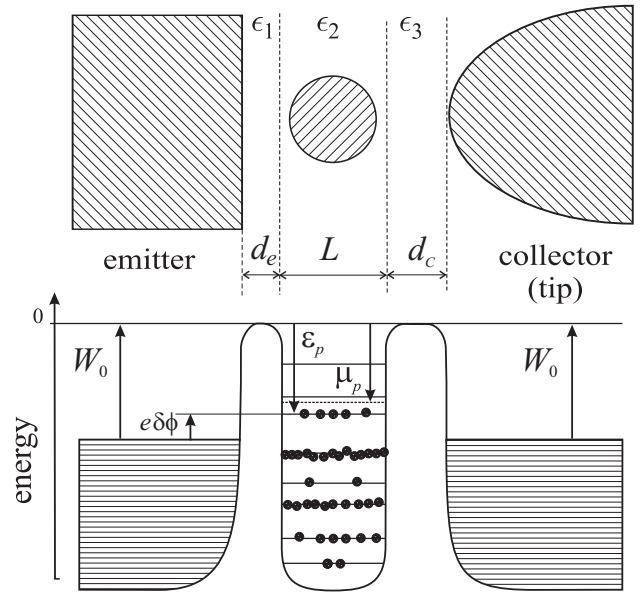}
\caption {The energy diagram for the structure Au/Au$_{40}$/Au
before application of voltage;
$\epsilon_{1}=\epsilon,\:\epsilon_{2}=\epsilon_{3}=1$ in experiments
\cite{Lih,Ohgi-2001,Ohgi,Ohgi-2004,Wang,Hou}.} \label{Fig1}
\end {figure}

In Refs. \cite {Ohgi-2001, Ohgi, Ohgi-2004} the structure with two
tunnel junctions (Fig. \ref {Fig1}) was represented
by a thick Au (111) film  covered by the $d_{e} \sim 10$ \AA \ thick
dielectric layer (with the dielectric constant $ \epsilon \approx
3$), on which the small spherical-like gold clusters were organized.
The tungsten tip (with small curvature of surface) of STM microscope
was covered by a gold film with $10^{3}$ \AA\: thickness. Therefore
we can consider all the three electrodes (two of them with a flat
surface) as being gold. Using the circuit approach of Ref.
\cite{Korotkov2}, the capacitances, tunnel resistances and a
``residual'' (fractional) charge $Q_{0}$ of granules were extracted
as fitting parameters from the measured dependence $I(V)$.

Earlier, a similar STM (Pt/Ir -tip) measurement for such a structure
($d_{e}\sim 1.4$ nm, $\epsilon \sim 2.7$) based on a gold islands
with monatomic height $H\approx 0.25 $ nm (disk-shaped clusters) was
carried out in the works \cite {Wang,Hou}.

The experiments demonstrated the following features of the $I(V)$
behavior:

1. The gap width of the zero conductance is approximately
proportional to the inverse radius of the disks (Fig. 4 in Ref.
\cite {Wang}) and spheres (Figs. 1(c) and 2(a) in Ref. \cite
{Ohgi-2001}). This does not allow one to establish unequivocally
classical or quantum origin of the gap. On the other hand, out of
the current gap, the steps of the staircase are clearly visible
(Fig. 3 in Ref. \cite {Wang} and Fig. 1(b) in Ref. \cite {Hou}).

2. For a disk, the gap width varies non-monotonically with
alteration of the collector--cluster distance under the fixed
emitter--cluster one (Fig. 3 in Ref. \cite {Hou}).

The possibility of a fractional charge at tunneling structures was
discussed in the Refs. \cite {17.,Azbel}. In the percolation systems
it is supposed that the charge $Q_{0}$ at the each granule has a
soliton origin. The value of this charge was calculated numerically
in Refs. \cite {17-1,17-2}.

However, not much attention was paid to this problem, as well as the
current gap origin on the $I-V$ curves and its asymmetry. Perhaps,
this problem is related to the fractional quantization (or
fractional statistics), when the decoupling of the spin and the
electron quantum numbers of a charge is important.

The aim of this work is the computation of the current-voltage
characteristic of the three-electrode structure, whose central
electrode is a metal cluster with different sizes and shapes.

The outline of this paper is as follows. In Sec.~II we formulate the
problem of the influence of the well-separated energy levels and
charging energy on the resonant tunneling through a metal cluster,
which is weakly coupled to two massive electrodes. Our main
assumption is that the thermal energy exceeds the width of
transmission resonance. The present approach describes the case of
strong inelastic scattering of electrons in the clusters that
corresponds to the full thermalization regime. We calculate electron
spectrum for the wells of various shape, chemical electron potential
and residual charge. This charge is a result of equilibrium with
electron sea. We use the density-functional approach for
determination of the total energy of metal cluster in an external
electric field. Next, we estimate the critical surplus charge of the
spheroidal cluster, which leads to the Coulomb instability for a
nonzero applied voltage. In Sec.~III, we consider electron transport
through a metallic quantum dot that can be described by a means of
master equation. The contact potential difference is taken into
account. In Sec.~IV we perform the numerical analysis of the $I-V$
curves for various set of parameters. The current jumps are
investigated in details for the tunneling structures based on the
magic and non-magic clusters. The model allows one to determine the
size of current gap and its asymmetry on a voltage.

\section{Preliminary analysis and formulation of the problem}

We consider spherical gold clusters whose radii vary in the range
$R\simeq 7 - 14$ \AA, $R=N^{1/3}_{0}r_{s}\Rightarrow N_{0} \simeq
100 - 600$ ($r_{s} = 3.01\:a _ {0}$ is the electron density
parameter, $a _ {0} $ is the Bohr radius). Similarly for disks of
monatomic thickness: $R\simeq 5 - 42.5$ \AA $\Rightarrow N_{0}\simeq
14 - 10^{3}$. We introduce  the characteristic charging energy
$\widetilde{E}_{C}=e^{2}/C$, where $C $ is a cluster capacitance
\cite {PHR-2005}. For spheres and disks we obtain
$\widetilde{E}_{C}\simeq 1.82 - 1.06$ and $3.2 - 0.42$ eV,
respectively. Temperatures of structures are $T <30$ K.

Let's determine the electron spectrum in spherical and cylindrical
wells (see Appendix A). For the above mentioned sizes, calculation
in both cases gives the close values of the spectrum discreteness,
$\Delta\varepsilon_{p}\approx 1.2 - 0.3$ eV, nearby to highest
occupied level $\varepsilon^{\rm HO}$ at $T=0$. Thus, for the whole
range of $R$ in experiments \cite {Ohgi-2001, Ohgi, Ohgi-2004, Wang}
we have to deal with a set of open 0D systems (quantum dots) . The
resulting inequality,
\begin {equation}
\widetilde{E}_{C}\approx\Delta\varepsilon_{p}\gg k_{\rm B}T, \label
{set-0}
\end {equation}
corresponds, apparently, to two coexisting structures at $I-V$
curves: effects of the spectrum quantization and the Coulomb
blockade. The current-voltage characteristic should represent a
superposition \emph{quantum} staircase with a step
$\sim\Delta\varepsilon_{p}/e$ and the \emph{classical}  Coulomb one
of the electrostatic nature with a step $\sim\tilde{E}_{C}/e$ along
a voltage. However, detailed measurements \cite{Ralph,Ohgi-2001,
Ohgi, Ohgi-2004,Gub-2002} performed to date do not yield an
unequivocal conclusion about the effect of electron quantization
levels upon the $I(V)$.

In our opinion, the discreteness of the spectrum actually determines
the zero conductance gap of the $I-V$ curves observed in \cite
{Ohgi-2001, Ohgi, Ohgi-2004, Wang, Hou}. Let's consider the problem
step by step.

\subsection{Structure in the absence of voltage}

The left and right electrodes (emitter and collector) represent the
electron reservoirs. Each  reservoir is taken to be in thermal
equilibrium. A continuum of states is assumed in reservoirs,
occupied according to the Fermi-Dirac distribution
\begin {equation}
f(\varepsilon^{e,c}+W_{0}^{e,c})=\left\{1+\exp[(\varepsilon^{e,c}
+W_{0}^{e,c})/k_{\rm B}T]\right\}^{-1}, \label{set-1}
\end {equation}
where $W_{0}>0$ is the work function for a semi-infinite metal. In
all cases energy $U_{0}<\varepsilon <0$ is counted off from the
vacuum level, $U_{0}<0 $ is the position of conductivity band of a
semi-infinite metal.

The electron chemical potential $\mu$ of a cluster in a quantum case
can be defined by the normalization condition
\begin {equation}
\sum_{p=1}^{\infty}f(\varepsilon_{p}-\mu)=N, \label {set-1a}
\end {equation}
where sum runs over all one-electron states, $N $ is the total
number of  thermalized conduction electrons in a cluster (taking
into consideration the surplus and lacking  electrons), and
\begin {equation}
f(\varepsilon_{p}-\mu)=\left\{1+\exp[(\varepsilon_{p}-\mu)/k_{\rm
B}T]\right\}^{-1}. \label {set-54}
\end {equation}

\begin {figure} [! t! b! p]
\centering
\includegraphics [width =.49\textwidth] {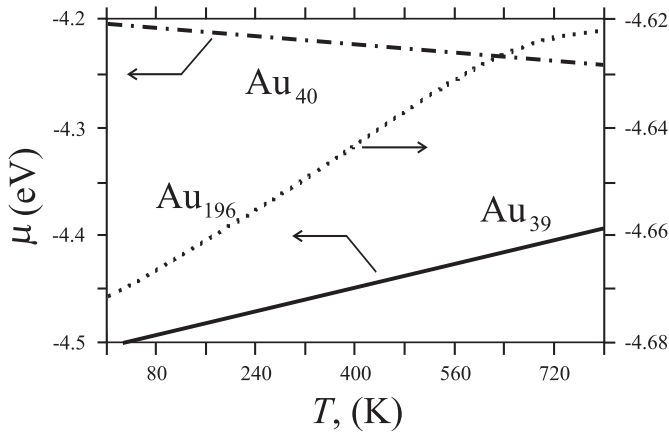}
\caption {Temperature dependence of chemical potentials of the
spherical gold  non-magic Au$_{39}$ and magic Au$_{40}$ and
Au$_{196}$ clusters.} \label {Fig2}
\end {figure}

If the electron spectrum is known, from Eq. (\ref {set-1a}) it is
possible to calculate $\mu$ of  Au$_{N_{0}}$, where $N_{0}$ is the
number of the conduction electrons of neutral clusters. The Fermi
level of non-magic clusters coincides with a real level in a
cluster. For the magic ones it lies between the energy terms. Figure
\ref {Fig2} depicts the chemical potential of some spherical
clusters as a function of temperature. Predictably, the dependence
is slack and
is completely determined by the level hierarchy in dots, and also by
the number of electrons. Calculations show, that the temperature
gradient of chemical potential can be both the positive and the
negative, and at some temperatures it can change a sign. Similar
behavior $\mu(N_{0},T)$ for magic clusters Na$_{N}$ have been
reported in Ref. \cite {75}.

A contact potential difference appears between a cluster and
electrodes (Fig. \ref{Fig1})
\begin {equation}
\delta \phi = (W _ {0} + \mu)/e. \label {set-1c}
\end {equation}
An equilibrium is reached by the noticeable charging of a cluster
since it capacitance is finite. If $|\mu|<W_{0}$, a cluster is
charged positively  by a charge $Q_{0}=-e(N^{\prime}-N_{0})> 0$,
where $N^{\prime}$ is determined by the solution of the Eq.
(\ref{set-1a}) with replacement $\mu\rightarrow -W_{0}$ for a
spectrum $\{\varepsilon_{p}\}$ shifted on $-e\delta \phi$, according
to the Koopmans' theorem \cite {PHR-2005}. In similar theories, the
presence of a contact potential difference were neglected.

Thus, we have
\begin {equation}
Q_{0}=C\delta \phi. \label {set-1b}
\end {equation}
This expression differs from the definition of $Q_{0}$ in the
orthodox theory (Eq. (29 b) in Ref. \cite {Averin}).

A quasi-classical approximation, $W_{0}+\mu(R)=\mu_{1}/R$, $
\mu_{1}=1.9$ eV$\times a_{0}$ \cite{PHR-2005}, gives $Q_{0}\simeq
+0.07\:e$. In a quantum case, it is necessary to make a replacement
$\mu\rightarrow \varepsilon^{\rm HO}$. Otherwise, when $|\mu|>W_{0}$
the cluster is charged negatively ($\mu\rightarrow \varepsilon^{\rm
LU}$).

The non-integer value of $Q_{0}$ (in terms of the elementary charge)
is due to the fact that the electron wave functions are not well
localized. Therefore, electrons cannot be interpreted as classical
particles, and a fraction of an electron (and its charge) can be
found in the other electrode \cite {17., Azbel}.

Value $Q_{0}\approx +0.5\, e$ better than the others corresponds to
the Kuzmin's and Likharev's experiment (Fig. 2(b) in Ref. \cite
{Lih}) in which the $I-V$ was measured  for the structure of two
electrodes (alloy Pb/Au) and granule In of radius $R\approx 100$ nm
separated by the  oxide films. It is interesting to estimate $Q_{0}$
by Eq. (\ref{set-1b}). As the work function of the alloy is unknown,
using accordingly 4 and 3.8 eV for Pb and In, we
 obtain considerably distinguished quantity $Q_{0}\approx +13.6\,
e$. However, assuming the exact quantity $Q _ {0}\approx +0.5\, e$,
it is possible to solve the inverse problem and to find a work
function of the alloy: 3.8012 eV (instead of 4 eV for Pb).

\subsection{Structure under  voltage $V$}

Between the emitter ($V=0$) and the collector the positive voltage
$V$ is applied. We consider a central electrode--granule in an
external electric field. In a weak electric field approach we
assume, that the ionic subsystem of a granule is not deformed, and
the electronic ``cloud'', generated by the own valence electrons, is
deformed only.

The total energy of a granule is the functional of nonhomogeneous
electron concentration, $\tilde{E}[n(\textbf {r})]$. The functional
contains a contribution responsible for the interaction of electrons
and ions with an external field,
\begin {equation}
e\int [n(\textbf {r})-n_{i}(r)](\textbf{E}\cdot\textbf{r})\, d^{
3}r. \label {set-2}
\end {equation}
For simplicity, we suppose that the ion distribution $n_{i}(r)$ is
spherically symmetric.

Let's write down an electron distribution of a granule as
\begin {equation}
n(\textbf{r})=n_{0}(r)+\delta n_{1}(r)+\delta n_{ 2}(\textbf {r}).
\label {set-3}
\end {equation}
Here, $n_{0}(r)$ is the electron density of neutral cluster in the
absence of the external field,
$$
\int n_{0}(r)\, d^{3}r = N_{0},
$$
$\delta n_{1}$ is the perturbation arising from the charging of
granule, and $\delta n_{2}(\textbf{r})$ is the next perturbation
arising from the external field which responses for the polarization
of a neutral granule,
\begin {equation}
\int\delta n_{1}(r)\,d^{3}r =\Delta N,\quad \int\delta n_{
2}(\textbf {r})\,d^{3}r=0, \label {set-4}
\end {equation}
where $\Delta N > 0$ and $\Delta N < 0$ correspond to negatively and
positively charged granule, respectively. We assume, that functions
$n_{0}(r)$ and $n_{1}(r)$ are spherically symmetrical, and
$n_{2}(\textbf{r})$ is axially symmetrical. Then one can expand the
$\tilde{E}[n(\textbf {r})]$ in the functional Taylor series down to
the second order of smallness with respect to $\delta n_{1}$ and
$\delta n_{2}$,
\begin {multline}
\tilde {E} [n (\textbf {r})] = \tilde {E} [n _ {0} (\textbf {r})] +
\sum _ { j} \int\frac {\delta \tilde {E}} {\delta
n (\textbf {r})} \, \delta n _ {j} \, d ^ {3} r + \\
\frac {1} {4} \sum _ {j, k} \iint\frac {\delta ^ {2} \tilde {E}}
{\delta n _ { j} (\textbf {r}) \delta n _ {k} (\textbf {r} ^
{\prime})} \, \delta n _ { j} (\textbf {r}) \, \delta n _ {k}
(\textbf {r} ^ {\prime}) \, d ^ {3} r \, d ^ { 3} r ^ {\prime} +
\ldots \label {set-5}
\end {multline}
Here the functional derivatives are taken at
$n(\textbf{r})=n_{0}(r)$, and indexes $j$ and $k$ runs 1 and 2
according to the definition (\ref{set-3}). The zeroth-order
expression $\tilde{E}[n_{0}(\textbf{r})]\equiv\tilde{E}_{00}$ is a
total energy of a cluster before the charging ($\Delta N=0$) and in
the absence of the external field ($E=0$). The functional derivative
\begin{equation}
\delta\tilde{E}/\delta n(\textbf{r})=\mu + e(\textbf{E}
\cdot\textbf{r}), \label {set-6}
\end{equation}
where the constant $\mu$ is the chemical potential of electrons of a
granule. For large clusters, $-\mu=W$ in the absence of charging and
the external field.

Finally, in the semiclassical approximation, we get
\begin{equation}
\tilde{E} = \tilde{E}_{00} + \mu\,\Delta N-e\Delta N \,\eta V +
(\Delta N)^{2}\widetilde{E}_{C}/2-\alpha E^{2}/2 \label{set-10}
\end{equation}
(see Appendix B).

Solving separately the electrostatic problem for the same structure
with fraction of the voltage $V>0$ (Fig. \ref{Fig1}), we obtain
\begin{equation}
\eta=\frac{\epsilon_{2} \epsilon_{3}(d_{\rm
e}+\epsilon_{1}L/2\epsilon_{2})}{\epsilon_{1}\epsilon_{2}d_{\rm
c}+\epsilon_{1}\epsilon_{3}L+\epsilon_{2}\epsilon_{3}d_{\rm
e}}\equiv \eta^{+}.\label{set-10a}
\end{equation}
Here, $L\equiv 2R,\:H$ for a sphere of radius $R$ and a disk of
thickness $H$, respectively, $\epsilon_{1}\equiv
\epsilon,\:\epsilon_{2}=\epsilon_{3}=1$. Under Eq. (\ref {set-10a})
one can find the values $ \eta^{+} \lesssim 0.65$ and
$\eta^{+}\lesssim 0.55$ in experiments \cite { Ohgi-2001, Ohgi,
Ohgi-2004,Wang, Hou} for the spherical-like and disk--shaped
clusters, accordingly.

Now we examine the problem of critical surplus charges of a cluster
in the presence of an external voltage. For convenience, further we
use a contraction $n\equiv\Delta N $.

\subsection{Coulomb instability of a cluster in electric field}

It is necessary to note, that even the vanishing external electric
field leads to the  instability of a cluster because of the
possibility of tunnelling of electrons. We assume, that a cluster
relaxed  in a metastable state over a period of time which is much
smaller, than that between acts of tunnelling. As a result of the
charging, the intrinsic mechanical ``stress'' leads to the ``Coulomb
explosion''. This problem was described in Ref. \cite {PHR-2005} for
a single spherical cluster in absence of the field. Extending these
results, one can write the following expression
\begin {equation}
\mp\left\{(-\mu_{e, i} + |e\eta V|)R/e + e/2\right\} \label{set-100}
\end {equation}
for the critical electronic or ionic charge in quasi-classical
approximation. For the range $V=$(0$,\: 2$V) we have:

i) $\eta \ll 1$. Transitions of electrons between the emitter and
the cluster occur more often, than between the cluster and the
collector, therefore the electrons are accumulated on the cluster.
In this case their maximal number is
$$
n _ {\rm max} \simeq W_{e}\:R/e^{2} + 1/2,
$$
where $W_{e}=W_{e0}-\mu_{e1}/R$ and $n_{\rm max}\simeq +2.5 - +6.5$.

ii) $\eta \approx 1$. Transitions of electrons between the cluster
and the collector occur more often, than between the cluster and the
emitter, therefore on the cluster the deficiency of electrons is
observed. Using the ion work function, this number determines as
$$
n_{\rm min} = -(W_{i} + |e\eta V|)\:R/e^{2}-1/2,
$$
where $W_{i} =W_{i0}-\mu_{i1}/R$ and $n_{\rm min}\approx -4 - -11$.
Similarly, for $V=$($-2,\:0$V)  we have:

i) $n_{\rm min}\simeq -W_{i}\:R/e^{2}-1/2 =-3.8 - -10.6.$

ii) $n_{\rm max} = (W_{e} + |e\eta V|)R/e^{2} + 1/2\approx +3 - +8.$

Below, the \emph{whole} numbers $[n_{\rm max}]$ and $[n_{\rm min}]$
bound the summation in (\ref{set-60}). The effect of spectrum
quantization can change these numbers no more than in $\pm 1$
according to the first inequality in (\ref {set-0}) (see Ref. \cite
{PHR-2005}).

Effective collision frequency of excited electrons in a cluster is
defined as \cite {15ao}
\begin {equation}
\tau_{\varepsilon}^{-1} = \tau^{-1}+v_{\rm F}R^{-1}, \quad v_{\rm F}
= (\hbar/mr_{s})(9\pi/4)^{1/3}, \label {setK-63}
\end {equation}
where $\tau$ is a relaxation time in the bulk of the metal, caused
by electron-electron collisions ($\tau\times 10^{14}=6.23$ s for Au
at $T=75$ K  \cite {17ao}), and $v_{\rm F}$ is the electron velocity
at the Fermi surface in the bulk. The estimation performed in Ref.
\cite {LetJTF-2000} gives a preferred electron collision on walls of
a dot, therefore $\tau_{\varepsilon}\simeq R/v_{\rm F}$. It leads to
$\tau_{\varepsilon}\Delta\varepsilon\simeq (0.52 - 0.17)\,\hbar$,
i.e. to a broadening of levels.

During the resonant tunneling, the discreteness of the spectrum can
be revealed only at low temperature (second of inequalities (\ref
{set-0})).

The electron thermalization occurs much faster than acts of
tunnelling. ``New '' electrons fill up a number of own electrons,
changing their distribution and, accordingly, the chemical
potential. This state of the cluster will be a starting state for
the next act of tunneling.

\section{Basic energy and kinetic relations}

We assume, that the \emph{total} energy of all three electrodes
$\tilde{E}$ does not change during the tunneling. In the case of
transition of $\delta N$ electrons from the emitter to the granule
(containing $n$ ``surplus'' electrons), we have from (\ref{set-10})
\begin{multline}
\delta\tilde{E}=-\delta N\:\overrightarrow {\varepsilon^{ e}}+
\delta N\:\varepsilon_{p}\\ + \frac {(-e)^{2}}{2C}[(n +\delta N)^
{2}-n^{2}]-e\delta N\eta^{+} V=0. \label {set-50bbb}
\end {multline}

In this equation we take into account that the electron ionizes from
the level $\overrightarrow{\varepsilon^{e}}$ on the emitter (whose
capacitance is equal to zero) and then sticks to the level
$\varepsilon_{p}$ in a granule with capacitance $C$.

By analogy with Ref. \cite{Been}, using Eqs. (\ref {set-50bbb}) and
(\ref {set-10}) for $\delta N=1$ , and than taking into account a
contact potential difference (\ref {set-1c}), for emitter--granule
transition we have
\begin {equation}
\overrightarrow {\varepsilon^{e}}=\epsilon_{p} + \tilde{E}_{C}
(n+1/2)-e\eta^{+} V, \label {set-50}
\end {equation}
where $\epsilon_{p} \equiv\varepsilon_{p}-e\delta \phi$. The arrow
on the top indicates the energies which are determined by transfers
according to Fig. 1. We suppose, that $n\equiv n(V)$ and $n=0$ at
$V=0$. However, the granule is charged by a $Q_{0}$ before voltage
applied. Therefore, we assume that $n$ is the result of the applied
voltage only.

For granule$-$emitter transition we have
\begin {equation}
\overleftarrow {\varepsilon ^ {e}} = \epsilon _ {p} + \tilde {E} _ {
C} (n-1/2)-e\eta^{+} V. \label {set-51}
\end {equation}
Similarly, for the granule$-$collector  and collector$-$granule
transitions we have
\begin {equation}
\overrightarrow{\overleftarrow {\varepsilon^ {c}}} = \epsilon_{ p} +
\tilde{E}_{C}(n\mp 1/2)+e(1-\eta^{+})V. \label {set-52}
\end {equation}
Here the upper/under arrows at the left correspond to the following
signs on the right. Independently of $n$ the relation
$$
\overrightarrow{\varepsilon^{e}}-\overleftarrow{\varepsilon^{ e}} =
\tilde{E}_{C}=\overleftarrow{\varepsilon^{ c}}-\overrightarrow
{\varepsilon^{c}}
$$
takes place. It agrees with well-known quasiclassical relation for
clusters, IP $-$ EA = $\tilde{E}_{C}$ (see, e.g. Ref. \cite
{PHR-2005}). Eqs. (\ref {set-50}) -- (\ref {set-52}) represent  a
``gold rule'' for transitions.

The tunneling of a single electron through barriers is determined by
the tunnel rates $\Gamma^{e,c}$ which depend on the junction
geometry and the voltage fraction $\eta$. In general, their
evaluation is far from a trivial problem \cite {Ralph, Azbel}. We
assume that they are small and the temperature is not too low, i.e.
\begin {equation}
k _{\rm B} T> \hbar (\Gamma ^ {e} + \Gamma ^ {c}) \ll \min
\{\Delta\varepsilon _ {p}, \:\tilde {E} _ {C} \}. \label {set-48}
\end{equation}

By analogy with the theory of Ref. \cite {Averin}, we introduce the
partial tunneling rates from electrodes to a granule
\begin {equation}
\overrightarrow{w_{n}^{e}} =2\sum_{ p}\Gamma (\overrightarrow
{\varepsilon^{e}})\, f(\overrightarrow{\varepsilon^{e}} +W_{V}^{
e})\, [1-f(\overrightarrow{\varepsilon^{ e}}-\overrightarrow{\mu _
{C}^{e}})], \label {set-55}
\end {equation}
\begin {equation}
\overleftarrow {w_{n}^{c}}=2\sum_{ p}\Gamma(\overleftarrow
{\varepsilon^{c}})\, f(\overleftarrow{\varepsilon^{c}} +W_{V}^
{c})\, [1-f(\overleftarrow{\varepsilon^{c}}- \overleftarrow{\mu _
{C}^{c}})], \label {set-55a}
\end {equation}
and from a granule to the electrodes
\begin{equation}
\overleftarrow{w_{ n}^{
e}}=2\sum_{p}\Gamma(\overleftarrow{\varepsilon^{ e}})\,
[1-f(\overleftarrow{\varepsilon^{
e}}+W_{V}^{e})]\,f(\overleftarrow{\varepsilon^{
e}}-\overleftarrow{\mu_{C}^{ e}}),
\end{equation}
\begin{equation}
\overrightarrow{w_{ n}^{c}}=2\sum_{
p}\Gamma(\overrightarrow{\varepsilon^{c}})\,
[1-f(\overrightarrow{\varepsilon^{c}}+W_{V}^{c})]\,
f(\overrightarrow{\varepsilon^{c}}-\overrightarrow{\mu_{C}^{c}}),
\label{set-56a}
\end{equation}
where the factor 2 takes into account the spin degeneration of
levels in electrodes. In view of the applied voltage (and charging
for a granule) the spectrums (see Eqs. (\ref {set-50}) - (\ref
{set-52})) and the chemical potentials are shifted in distributions
(\ref {set-1}) and (\ref {set-54})
$$
W_{V}^{e}\equiv W_{0}^{e},\quad
\overleftarrow{\overrightarrow{\mu_{C}^{e}}}=\mu-e\delta
\phi+\tilde{E}_{ C}(n \mp 1/2)-e\eta^{+} V,
$$
$$
\overleftarrow{\overrightarrow{\mu_{C}^{c}}}=\mu-e\delta
\phi+\tilde{E}_{ C}(n\pm 1/2)+e(1-\eta^{+}) V, \quad W_{V}^{c}=
W_{0}^{c}+eV.
$$
As the first approximation of the perturbation theory \cite
{PHR-2005}, for small $V$, $\mu$ is determined  not only by the
formal shift of the well depth, but also by the number of conduction
electrons in the state ($N=N_{0}+n_{\rm q}$, $n_{\rm
q}=n+[Q_{0}]/e$). The use of the chemical potentials is correct in a
quasi-equilibrium state, i.e. when the intervals between acts of
tunneling are much longer than the relaxation time of a granule. It
is also supposed, that the external field and the Coulomb blockade
do not remove degeneration of levels. In this case, at $T=0 $,
probabilities (\ref {set-55}) -- (\ref {set-56a}) will be nonzero
for the intervals:
$$
\overrightarrow{\mu_{C}^{e}}\leq\overrightarrow{\varepsilon^{e}}\leq
-W_{V}^{e}\leq\overleftarrow{\varepsilon^{e}}\leq\overleftarrow{\mu_{C}^{e}},
$$
$$
\overleftarrow{\mu_{C}^{c}}\leq\overleftarrow{\varepsilon^{c}}\leq
-W_{V}^{c}\leq\overrightarrow{\varepsilon^{c}}\leq\overrightarrow{\mu_{C}^{c}}.
$$

Let's denote the total  electron transition rates to a granule and
back on electrodes, as
$$
w_{n}^{\rm in}=\overrightarrow{w_{n}^{ e}}+\overleftarrow{w_{
n}^{c}},\quad w_{ n}^{\rm out}=\overleftarrow{w_{
n}^{e}}+\overrightarrow{w_{n}^{c}}.
$$

In the limit of weak tunneling, the probability $P_{n}$ of the
finding of $n$ for the above mentioned electrons at central
electrode is defined by the master equation \cite {Averin}
\begin{equation}
\dot{P_{n}}=w_{n+1}^{\rm out}\,P_{n+1}+w_{ n-1}^{\rm
in}\,P_{n-1}-(w_{n}^{\rm in}+w_{ n}^{\rm out})\,P_{ n}.
\label{set-57}
\end{equation}
The requirement of the stationarity, $\dot{P_{n}}=0$, gives  the
recurrent relation
\begin{equation}
P_{n+1}=P_{ n}\,w_{n}^{\rm in}/w_{n+1}^{\rm out}. \label{set-58}
\end{equation}

The dc current flowing through a quantum granule (with restriction
on its instability (\ref {set-100})), is determined as
\begin{equation}
I=-e\sum_{n_{\rm min}<0}^{n_{\rm max}>0} P_{
n}\left(\overrightarrow{w_{ n}^{ e}}-\overleftarrow{w_{ n}^{
e}}\right)=- e\sum_{n_{\rm min}<0}^{n_{\rm max}>0} P_{
n}\left(\overrightarrow{w_{ n}^{c}}-\overleftarrow{w_{ n}^{
c}}\right). \label{set-60}
\end{equation}

Let's consider an exotic case of the ``strong quantization'' for
electron spectrum \cite {Averin}:
$$
\Delta \varepsilon_{ p}\gg \tilde{E}_{ C}.
$$
This regime is hypothetically reached by a significant increase of
the cluster capacitance (the cluster shape must be changed to the
needle-like or disk-like one under the condition that its volume is
fixed (see, e.g. Ref. \cite {PHR-2005}, experiments with
 disks \cite {Wang} and  nanowires \cite {Millo})). Thus the
residual charge $Q_{0}$ in (\ref {set-1b}), which provides a contact
potential difference, is proportional to the capacitance and can
have large magnitude. When the voltage is applied, the charge, which
is caused by the transferring surplus electrons, is much less than
$Q_{0}$. Therefore it insignificantly influences the cluster
energetics. In reality, the inequality $\Delta\varepsilon_{p}\gg
\tilde{E}_{C}$ is not possible for the atomic chain \cite
{PHR-2005}. Nevertheless, this case is useful from the methodical
point of view to analyze the current gap of $I-V$ characteristics.

As an assumption, we use the fixed tunnel rates at the Fermi level
in the emitter. It is correct for the small voltages, $eV\ll W$.
Neglecting in (\ref{set-50}) -- (\ref{set-52}) terms
$\sim\tilde{E}_{C}$, it is easy to obtain the result, similar to
Ref. \cite {Averin}:
\begin{equation}
I=I_{0}\sum_{p}\left[f(\varepsilon^{ e}+W_{0}^{e})-f(\varepsilon^{
c}+W_{V}^{c})\right], \label{set-63}
\end{equation}
where $I_{0}=2e\Gamma^{e}\Gamma^{c}/(\Gamma^{e} + \Gamma^{c})$. At
$T\longrightarrow 0$, it is convenient to write Eq. (\ref {set-63})
as a "combination" of the staircase functions:
\begin{equation}
I/I_{ 0}\longrightarrow \sum_{ p}
\left[\theta(\xi_{p}+W_{0})-\theta(\zeta_{p}+W_{V}^{c})\right],
\label{set-63a}
\end{equation}
where for shifted spectrums in the emitter and the collector the
folowing notations are introduced:
$\xi_{p}\equiv\epsilon_{p}-e\eta^{+} V$ and $\zeta_{p} =
\epsilon_{p}+e(1-\eta^{+})V$.

Let's remind, that the expressions in this section are written down
for $V>0$. In the case $V<0$, the $I(V)$ can be received, if we set
$V=0 $ on a collector  and $V>0$ on the emitter. Now for Eqs.(\ref
{set-10a}), (\ref {set-50})--(\ref {set-52}) we have
\begin{equation}
\eta^{-}=\frac{\epsilon_{2} \epsilon_{3}(\epsilon_{1}d_{\rm
c}/\epsilon_{3}+\epsilon_{1}L/2\epsilon_{2})}
{\epsilon_{1}\epsilon_{2}d_{\rm c}+\epsilon_{1}\epsilon_{3}L+
\epsilon_{2}\epsilon_{3}d_{\rm e}}, \label{set-63ab}
\end{equation}
$$
\overleftarrow{\overrightarrow{\varepsilon^{c}}}= \epsilon_{
p}+\tilde{E}_{ C}(n\pm 1/2)-e(1-\eta^{-}) V,
$$
$$
\overleftarrow{\overrightarrow{\varepsilon^{e}}}= \epsilon_{
p}+\tilde{E}_{ C}(n\mp 1/2)+e\eta^{-} V.
$$
Also using
$$
W_{V}^{c}\equiv W_{0}^{c},\quad
\overleftarrow{\overrightarrow{\mu_{C}^{c}}}=\mu-e\delta
\phi+\tilde{E}_{ C}(n \pm 1/2)-e(1-\eta^{-}) V,
$$
$$
\overleftarrow{\overrightarrow{\mu_{C}^{e}}}=\mu-e\delta
\phi+\tilde{E}_{ C}(n\mp 1/2)+e\eta^{-} V, \quad W_{V}^{e}=
W_{0}^{e}+eV
$$
and $I \rightarrow -I$, for derived dependence $I(V)$ at $V>0$ it is
necessary to make mirror reflection relative to $V=0 $ on area
$V<0$. In such a case, for example, value of the $I_{0}$ in Eq.
(\ref{set-63a}) on $V>0$ differs from $I_{0}$ on $V<0$ since the
tunnel rates are different.

In the general case, for calculation of $I-V$ (\ref{set-60}) it is
necessary to know probabilities $P_{n}$. Their statistical
determination is a complicated problem \cite {Brack-91}. In the
experiments, the size of the cluster and its location are known only
approximately, therefore detailed calculations of $P_{n}$ are not
suitable. Using the recurrent relations we can find the ratios
$P_{n\neq 0}/P_{0}$.

\section{APPLICATION AND DISCUSSION}

\begin {figure}
\centering
\includegraphics [width =.49\textwidth] {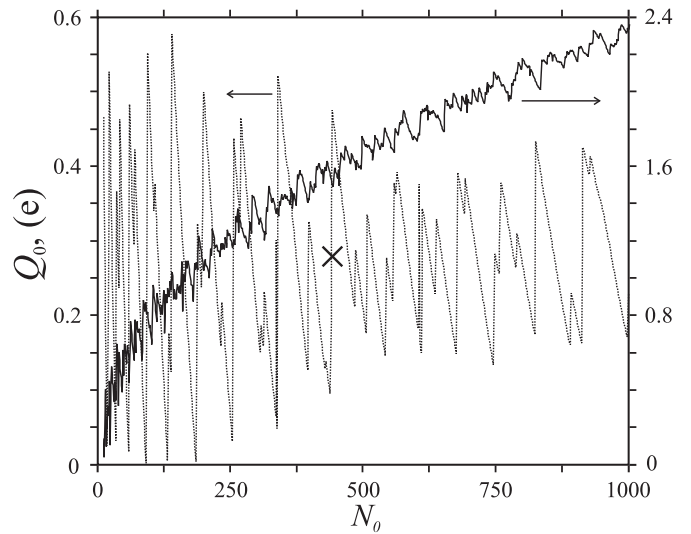}
\caption {The calculated size dependence of the residual charge
$Q_{0}$ (\ref {set-1b}) for the structure based on the clusters of
various shape: sphere (dotted line) and disk (solid line). For
illustration, $Q_{ 0}$ of magic sphere Au$_{439}$ is marked as
$\times$.} \label {Fig3}
\end {figure}

The cluster is charged positively before the application of voltage.
The size dependence of a charge $Q_{0}(N_{0})$ for referred gold
clusters is demonstrated on Fig. \ref {Fig3}.
For the above mentioned sizes of spherical clusters, $Q_{0}<e$.
However, $Q_{0}$ can accept a values larger then  $e$ for the disks
of monatomic thickness. Additional charging of the cluster can lead
to the Coulomb instability, because the quantity $Q_{0}$ is close to
a critical charge \cite {PHR-2005}.

Moreover, cluster's anomalous electrostriction is possible as a
result of the charging \cite{Pog-Solid}.

Setting the collector$-$granule distance $d_{c}$, parameter
$\beta=\Gamma^{e}/\Gamma^{c}$ and using the recurrent relation
(\ref{set-58}) for Eq. (\ref {set-60}), it is possible  to calculate
the reduced dc  current $\widetilde{I}\equiv I/(eP_{0}\Gamma^{e})$.
We do not  evaluate separately the threshold voltages, in our scheme
it appears automatically.

\begin {figure}
\centering
\includegraphics [width =.49\textwidth] {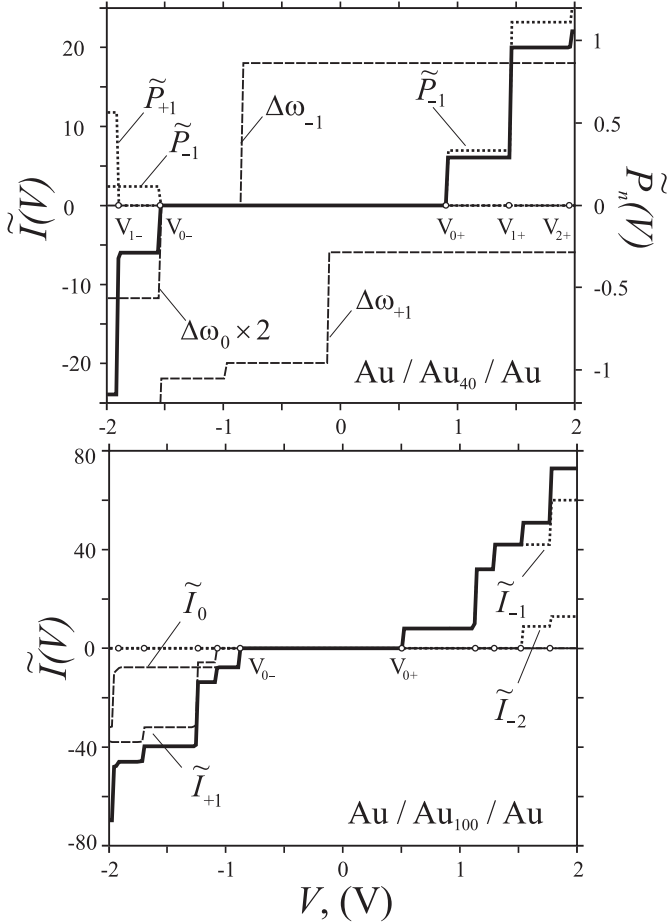}
\caption {The current--voltage curves (solid lines) and its
components, calculated from Eq. (\ref {set-60}) ($\beta = 1$,
$\eta^{+} = 0.1$, $T = 30$ K). $\Delta\omega_{n}(V)$ is given in
$\Gamma^{e}$ units.} \label {Fig4}
\end {figure}

The results of calculations of the $I-V$ characteristics for the
structures Au/Au$_{N_{0}}$/Au, based on spherical clusters, are
presented in Fig. \ref {Fig4}.
For completeness of analysis, the voltage behavior of the reduced
probabilities $\widetilde{P}_{n}(V)\equiv P_{n}/P_{0}$ and the
stream $\Delta \omega_{n} =
\overrightarrow{w_{n}^{e}}-\overleftarrow{w_{n}^{e}}$ are given
also.

The current jumps are  stipulated by the jumps of
$\widetilde{P}_{-1}(V)$ and $\Delta\omega_{0}(V)$, because the
current is formed by their product. Making use of the equality
$\widetilde{I} \equiv \sum_{n}\widetilde{I}_{n}(V)$ in accordance
with Eq. (\ref {set-60}) one can fix also the ``threshold'' values
$n$.

As one can surmise, the jump of probability $\widetilde{P}_{-1}(V)$
causes the current jump in the threshold voltage $V_{0+}$. Note that
the jumps of the stream $\Delta\omega_{0}$ at $V=V_{0+}$ and
$V_{0-}$ are determined by jumps of $\widetilde{P}_{-1}(V)$ or
$\Delta\omega_{0}$, respectively.

As is seen from Fig. 4, the role of partial current components (with
$|n|>1$) grow with increasing $N_{0}$. The charging leads to energy
shift of spectrum according to Eqs. (\ref{set-50}) -- (\ref
{set-52}). Thus the different parts of a spectrum are involved
during tunneling. The electron chemical potential of magic cluster
Au$_{40}$ does not coincide with a energy level at a zero voltage.

The current gap width $\Delta V_{g}=V_{0+}+|V_{0-}|$ in all cases is
determined by values $n=0,-1$, and the boundaries can be defined by
the position of lowest unoccupied level as
$$
|V_{0\pm}|=\frac{\tilde{E}_{ C}/2 +
\Delta\varepsilon}{(2-\eta^{\pm})},
$$
where $\Delta\varepsilon=\{\mu_{p}-\varepsilon^{\rm LU} \text{ and}
\:\: 0\}$ for the magic and non-magic clusters, accordingly. The
probability
 $P_{-1}$ prevails over $P_{+1}$, because the ``granule$-$collector'' stream
is more than a ``emitter$-$granule'' one and the granule is charged
positively (i.e. $n <0$).

\begin {figure}
\centering
\includegraphics [width =.49\textwidth] {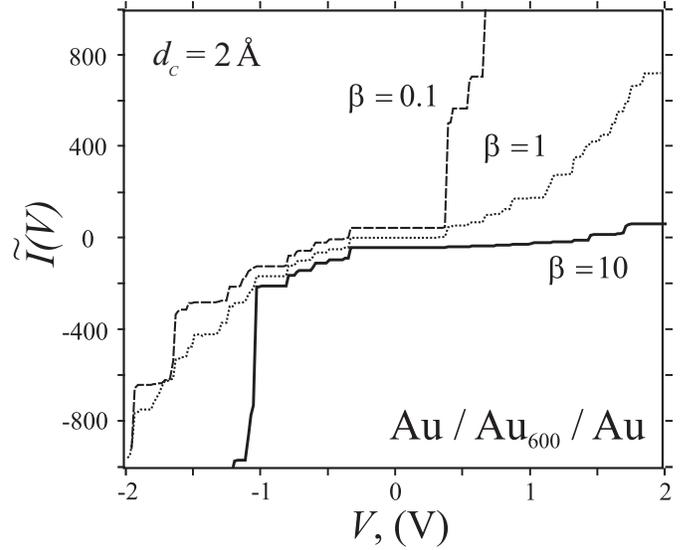}
\caption {Calculated $I-V$ curves at $T = 30$ K for structure based
on spherical clusters. For presentation the curves are shifted
slightly on a vertical.} \label {Fig5}
\end {figure}

Within the applied voltage the $I-V$ characteristics versus
$\eta^{+}$ are shifted to the right and the gap width decreases a
little. The calculated $I-V$ curves of the structure
Au/Au$_{600}$/Au for fixed $\eta^{+}$ and different $\beta$ are
shown in Fig. \ref{Fig5}. In this model, the current gap is
practically independent on $\beta$, however, the current jumps are
strongly dependent on the value of $\beta$, which, in its turn, has
no influence on threshold voltages.

\begin {figure}
\centering
\includegraphics [width =.49\textwidth] {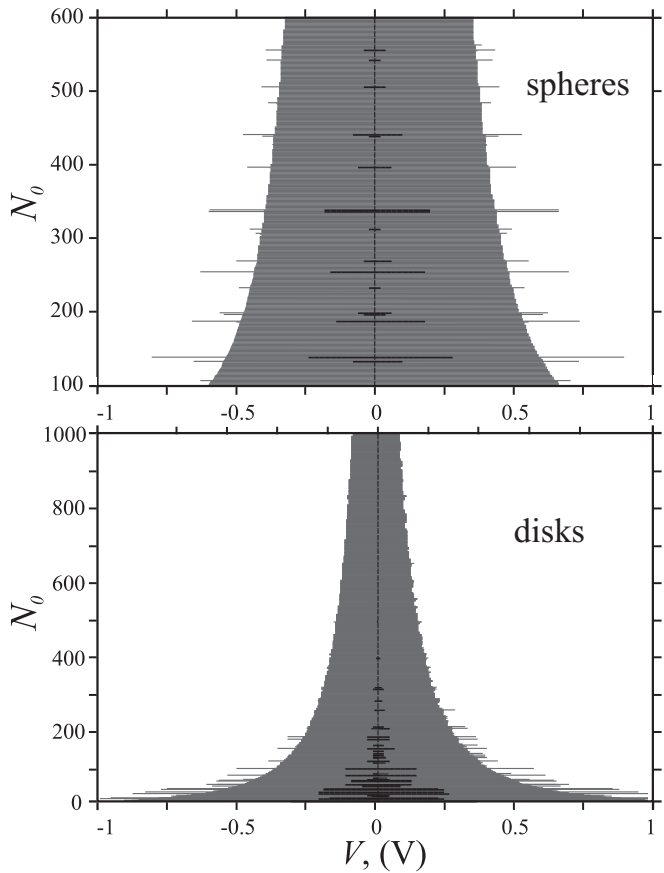}
\caption {The current gap calculated from Eq. (\ref {set-60})
($d_{c}=2$\:\AA and $\beta=10$). Solid lines show the gaps
calculated from Eq. (\ref{set-63}) for the case of ``strong
quantization''.} \label {Fig6}
\end {figure}

In order to illustrate our results, in Fig. \ref{Fig6}, we compare
the size dependences $\Delta V_{g}(N_{0})$ calculated from Eqs.
(\ref {set-60}) and (\ref{set-63}) for spheres and disks. The
largest quantities $\Delta V_{g}$ correspond to the magic granules,
for which $\Delta\varepsilon\neq 0$. For the case of ``strong
quantization'' the size of a current gap for non-magic clusters is
equal to zero explicitly, because the emitter Fermi level is in line
to the closed levels in cluster. Calculations demonstrate the
non-monotonic dependence $\Delta V_{g}(\eta)$. These results shows
also, that a charging leads to the growth of a gap.
\begin {figure}
\centering
\includegraphics [width =.49\textwidth] {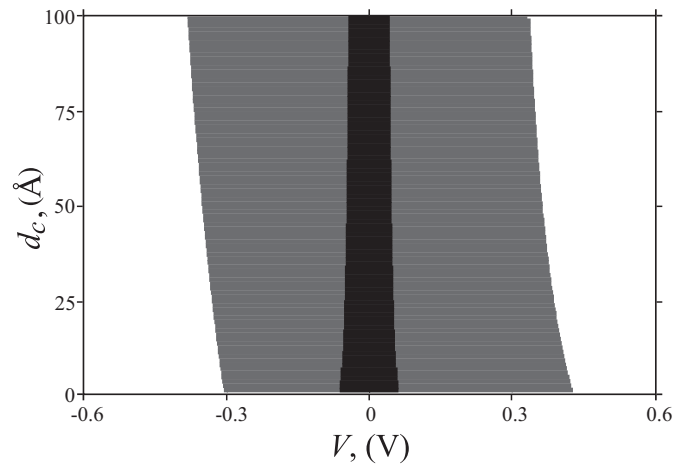}
\caption {Calculated current gaps ($\beta=10$) from Eqs. (\ref
{set-60}) and (\ref {set-63}) for magic disk Au$_{178}$ .}
\label{Fig7}
\end {figure}

The actual forms of current gap for the structure based on magic
disk Au$_{178}$   ($R\approx 3.5$ nm) are plotted in Fig. \ref
{Fig7}. Our calculations show that the dependence $\Delta
V_{g}(\eta)$) is slightly non-monotonic. However, in experiments
\cite {Wang, Hou} the gap varied as $0.8\rightarrow 0.4\rightarrow
0.7$ V for the cyclic variation of $d_{c}\approx 1\rightarrow
2\rightarrow 1$ $\mathrm{\mathring{A}}$. The reasons of such
considerable difference, apparently consist in the following
effects: broadenings of levels, amplifications nonlinearities  in
the strong electric field, and in energy dependence of tunneling
rates. At high rates the capacitance ceases to be classical and can
strongly grow ($\tilde{E}_{C}\rightarrow 0$) \cite {Wang2, Konig},
showing non-monotonic dependence from $\Gamma^{c}$. It means, that
in a reality we deal with the intermediate cases (between limiting
estimations from Eqs. (\ref {set-60}) and (\ref {set-63}) in Fig.
7). Only in these situations the observed behavior of gap width is
more or less clear.

Let's discuss other features of the tunnel construction. In spite of
the fact that the emitter and a collector are made of one material,
the chemical potentials of electrons are not equal to each other:
the emitter is represented by a thick film of Au (111), and a
collector is a polycrystal of Au. Their work functions are different
\cite{25b}. We assume also that $W <W_{0}$ for low-dimensional
systems \cite{PHR-2005}. Except for it the emitter is covered with a
dielectric film, that also influences a electron work function. We
can estimate this contribution.

Proceeding from indirect experimental measurements \cite{5}, the
work function decreases with growth dielectric constant $\epsilon$
of coating.  The calculations of the electron work function $W_{d}$
for cylindrical nanowires in a dielectric confinement are done in
Ref. \cite{6}: $W_{d}$ decreases approximately on 20\% at magnitude
as $\epsilon$ rise from 1 to 4. The basic contribution thus can be
related to the change  of electrostatic dipole barrier which
contribution to a work function of system gold$-$vacuum makes up to
30\% \cite {25b}. Hence, this contribution also makes a upper limit
of the change of $W_{d}$ for metal$-$dielectric$-$vacuum system.
Owing to $W_{d}<W_{0}$ the inequality $W>W_{d}$ is possible, that
leads to negative charging of the cluster before the application of
voltage. The film$-$cluster contact also will change the cluster
energetics. At last, the metal-nonmetal transition for gold cluster
can appears \cite {Boyen}.

\section{Summary}

In this paper, we presented an approach for the calculation for the
$I-V$ characteristics of SET based on the metal clusters.

In the framework of the particle-in-a-box model for the spherical
and disk-shaped gold clusters, the electron spectrum was calculated.
In this model, the work function of clusters is smaller that of
semi-infinite metal electrodes. It resulted in the appearance of a
contact potential difference between a cluster and electrodes.
Residual charge is equal to non-integer elementary charge $e$, which
corresponds to the charge of cluster in chemisorption regime. For
the small spherical clusters, positive charge is less than $e$.
However, this charge can accept a values larger than $e$ for the
disks of monoatomic thickness. Additional charging of the clusters
can lead to the Coulomb instability, because it is close to a
critical charge. The charging results in the energy shift of the
spectrum.

The current-voltage characteristics were analyzed taking into
account the contact potential difference. The approach was applied
to calculate the trapped offset charges that determine the transport
behavior of molecular-like structures. For single-electron
transistors based on the small gold clusters the current gap and its
voltage asymmetry  were computed. The largest quantities gap
correspond to the magic granules. For the case of ``strong
quantization'', the size of a current gap for non-magic clusters is
equal to zero explicitly, because the emitter Fermi level is in line
to the closed levels in cluster. The  results shows, that a charging
leads to the growth of a gap. The evaluations demonstrated a
slightly non-monotonic dependence (in comparison with the
experiments) of the gap size from cluster -- collector distance. At
high tunneling rates the capacitance ceases to be classical and must
strongly change.

\acknowledgments {We are grateful to W. V. Pogosov for reading the
manuscript. This work was supported by the Ministry of Education and
Science of Ukraine (Programme "Nanostructures") and Samsung
Corporation.}

\begin{appendix}

\section{Electron spectrum in cylindrical-like clusters}

As an approximation, the profile of the one-electron effective
potential in the cluster can be represented as a potential well the
depth $U_{\rm 0}<0$. The three-dimensional Schr\"{o}dinger equation
for a quantum box can be separated to the one-dimensional equations.
The spectrum of wave numbers in a spherical and cylindrical
potential wells are determined from the continuity condition of a
logarithmic derivative of the wave function on the boundaries.

For a disk of radius $R $ and thickness $H $ it is necessary to
solve the equation:
\begin{equation}
k_{nm}\frac{I_{m}^{\prime}(k_{nm}R)}{I_{m}(k_{nm}R)}=
\varkappa_{nm}\frac{K_{m}^{\prime}(\varkappa_{nm}R)}
{K_{m}(\varkappa_{nm}R)}. \label{set-01}
\end{equation}
Here $I_{m}$ is the Bessel function, $K_{m}$ is the McDonald
function, the stroke denotes a derivative over an argument,
$k_{nm}=\sqrt{k_{0}^{2}-\varkappa_{nm}^{2}}$, $\hbar
k_{0}=\sqrt{2m_{e}|U_{0}|}$, and $m_{e}$ is the electron mass. The
$n=1,2,3...$ number the roots of the Eq. (\ref{set-01}) for the
fixed $m=0, \pm 1, \pm 2... $.

Using the perturbation theory \cite {PHR-2005}, the finite well
problem can  be reduced to the infinite well one:
$$
k_{nm}=k_{nm}^{(0)}+k_{nm}^{(1)}+...,\quad
|k_{nm}^{(1)}/k_{nm}^{(0)}|\ll 1,
$$
where $k_{nm}^{(0)}$ defines the spectrum of infinitely depth well.
Numbers $k_{nm}^{(0)}$ are determined by solutions of the equation
$$
I_{m}(k_{nm}^{(0)}R)=0.
$$
In the first approximation we have
$$
k_{nm}^{(1)}=\frac{k_{nm}^{(0)}K_{m}(\varkappa_{nm}^{(0)}R)}
{R\varkappa_{nm}^{(0)}K_{m}^{\prime}(\varkappa_{nm}^{(0)}R)}.
$$
The estimation gives $k_{nm}^{(1)}<0.07k_{nm}^{(0)}$, confirming
sufficient accuracy of the theory.

Quantization of the wave vector $k_{s}$ along the cylinder axis is
determined by the solution of the equation:
$$
k_{s}H =s\pi -2\arcsin (k_{s}/k_{0}),
$$
where $s$ is the integer number. Neglecting the area near cylinder
edges, the energy spectrum is calculated by a simple way as follows
$$
\varepsilon_{nms}=U_{0}+\frac{\hbar^{2}}{2m_{e}}(k_{nm}^{2}+
k_{s}^{2}).
$$

In addition to the spin degeneration, there is a double degeneration
with respect to the sign of index $m$, since $k_{n,m} =k_{n,-m} $.
Further, the spectrum of cluster is denoted as $\varepsilon _ {p}$,
$p=1,2,3...$ is the number of \emph {an one-electron state}. All
levels are numbered in order to increase energies.

\section{Energy of cluster in external electrical field}

Using spherical coordinates, we remove the center point $z=0$ in a
center of a granule, and we direct a $z$ axis from a collector to
the emitter under the conservation the potential difference between
them. Then an electric field $\textbf{E}=E\hat{\textbf{z}}$, where
$\hat{\textbf{z}}$ is a unit vector along an $z$ axis.

As the surplus charge is effectively distributed over a surface, it
is quite reasonable to use
\begin {equation}
\delta n_{1}(r)=A\,\delta_{\rm D}(r-R), \label{set-7}
\end {equation}
where $\delta_{\rm D}(r-R)$ is the Dirac delta$-$function. In spite
of the fact that the form (\ref{set-7}) corresponds to the total
screening of a surplus charge inside a granule, this form
 is rather convenient for the calculations.
Then, we use the linear response approach (see, e.g. Ref. \cite{42})
\begin {equation}
\delta n_{2}(r,\theta)=Y (\textbf{r})\,E\cos\theta. \label{set-8}
\end {equation}
The constant $A$ in Eq. (\ref {set-7}) and spherically symmetric
function $Y(r)$ in (\ref {set-8}) are determined from the
normalization condition (\ref {set-4}) and a global minimum of the
functional, $\delta\tilde{E}[n(\textbf{r})]\rightarrow 0$.

One of the terms, interesting for us, is
$$
-e\int\delta n_{1}(r)\varphi(z)\,d^{3}r,
$$
where $\varphi$ is an external electrostatic potential. In the case
of $V>0$ and vacuum collector-emitter space $\varphi (z) =
V(z-d_{e}-L/2)/d$, where $d=d_{e}+L+d_{c}$.

After the integration in spherical coordinates, the term, which is
proportional to $z$, vanishes, and, as a result, we have $-e\Delta
N\eta V$, $\eta$ is  a fraction of a voltage. Other three terms
\begin{equation}
\int \frac{\delta n_{ 1}(\textbf{r})\delta n_{
2}(\textbf{r}^{\prime})+\delta n_{ 1}(\textbf{r})\delta n_{
1}(\textbf{r}^{\prime})+\delta n_{ 2}(\textbf{r})\delta n_{
2}(\textbf{r}^{\prime})}{|\textbf{r}-\textbf{r}^{\prime}|}\, d^{
3}r\:d^{ 3}r^{\prime}. \label{set-9}
\end{equation}
give a basic contribution to the second order of expansion (\ref
{set-5}). The first integral in Eq. (\ref{set-9}) for the functions
(\ref {set-7}) and (\ref {set-8}) vanishes after the integration on
corners, second and the third ones were calculated earlier at the
definition of the ionization potential  (see, e.g. Ref.
\cite{PHR-2005}) and polarizabilities of a cluster $\alpha
=-(4\pi/3) \int_{0}^{\infty}Y (r)r^{3}dr \equiv R_{\rm eff}^{3}$
\cite{42}. Finally we obtain Eq. (\ref{set-10}).
\end {appendix}

\end {document}